\documentclass[hyper,letterpaper,12pt]{article}
\usepackage{a4wide}
\usepackage{amsmath}
\usepackage[
      colorlinks=true,
      linkcolor=blue,
      urlcolor=blue,
      filecolor=black,
      citecolor=blue,
            ]{hyperref}
\usepackage{color}
\usepackage{graphicx}
\usepackage{amsfonts}
\usepackage{amssymb}
\usepackage{epsfig}
\usepackage{subfigure}
\usepackage{amsmath}
\usepackage{latexsym}
\usepackage{mathrsfs}

\begin{document}
\title{\bf \Large Holographic Van der Waals  phase transition for a hairy black hole}

\author{\large~~
Xiao-Xiong Zeng$^{1,2}$\footnote{E-mail: xxzeng@itp.ac.cn}
~~Yi-Wen Han$^3$\footnote{E-mail:  hanyw1965@163.com}~,
~
\date{\today}
\\
\\
\small $^1$ School of Material Science and Engineering, Chongqing Jiaotong University,\\
\small       Chongqing ~400074, China\\
\small $^2$State Key Laboratory of Theoretical Physics, Institute of Theoretical Physics,\\
\small Chinese Academy of Sciences, Beijing 100190,  China\\
\small $^3$ School of Computer Science and Information Engineering,\\
\small Chongqing Technology and Business University, Chongqing 400070, China\\}

\maketitle

\begin{abstract}
\normalsize The Van der Waals(VdW)  phase transition in a hairy black hole is investigated  by analogizing its charge, temperature, and entropy as the temperature, pressure, and volume in the fluid respectively. The two point correlation function(TCF), which is dual to the geodesic length,  is employed to  probe this phase transition. We find the phase structure in the temperature$-$geodesic length plane resembles as that in the temperature$-$thermal entropy plane besides the scale of the horizontal coordinate. In addition, we find
 the equal area law(EAL) for the first order phase transition
  and  critical exponent of the heat capacity for the second order phase transition in the temperature$-$geodesic length plane are  consistent with that in temperature$-$thermal entropy plane, which implies that the TCF  is a good probe to probe the phase structure of the back hole.
\end{abstract}

\newpage

%\tableofcontents

\section{Introduction}
Phase transition of the black holes is always a hot topic in theoretical physics for it provides a platform to relate the gravity, thermodynamics, and quantum theory. Phase transition in AdS space is more fascinated owing to the AdS/CFT correspondence. The Hawking-Page phase transition[1], which  portrays the transition of thermal gas to the Schwarzschild black hole, can be used to describe the confinement to deconfinement transition of the quark-gluon plasma in  Yang-Mills theory[2]. The phase transition of
a scalar field
condensation around
a  charged AdS black holes  can be used to describe the superconductor transition[3-5]. Especially one often can use the nonlocal observables such as holographic entanglement entropy, Wilson loop, and   TCF  to probe these phase transitions[6-7].

VdW phase transition is another important properties of the charged AdS black hole. It was observed that a charged black hole will undergo a first order phase transition, and a second order phase transition successively  as the charge of the black hole increases from small to large,  which is analogous to the van der Waals liquid-gas phase transition [8]. This phase transition was perfected recently by  regarding  the cosmological constant  as the  pressure  for in this case we need not any analogy[9-10].

Whether the VdW phase transition of the charged black hole can be probed by the nonlocal observables thus is worth exploring. Recently, Johnson[11]  investigated the  VdW phase transition of the Reissner-Nordstr\"om AdS black hole from the viewpoint of holography and found that the phase structure in the temperature-entanglement entropy plane resembles as that in temperature-thermal entropy plane. Thereafter Nguyen [12] investigated exclusively the EAL  in the temperature-entanglement entropy plane and found that the EAL   holds  regardless of the size of the entangling region.
Now
 there have been some extensive studies [13-19] and all the results show that as the case of thermal  entropy, the entanglement entropy  exhibits the similar VdW phase transition.

In this paper, we are going to use the equal time TCF  to probe the phase structure  of the hairy black holes.
It has been shown that the TCF   has the same effect as that of the holographic  entanglement entropy as it was used to probe the thermalization  behavior[20-33], thus it will be interesting to explore whether this observable can probe the phase structures of the black holes. The hairy AdS black hole is a solution of Einstein-Maxwell-$\Lambda$ theory conformally
coupled to a scalar field[34]. This model has at least two advantages. One  is that it is an  ordinary and tractable model
for studying superconducting phase transition with consideration of 
  the  back-reaction of the scalar field.
 Another
is that it  exhibits more fruitful phase transition behavior, namely not only the VdW behaviour in both the charged and uncharged cases, but also reentrant
phase transition in the charged case[35].  In this paper, we mainly concentrate on the VdW phase transition behaviour. Beside in the thermal entropy-temperature plane,  we will  also study the EAL    and  critical exponent of the heat capacity in the geodesic length-temperature plane. We  find  the results obtained in both framework  are consistent.

Our paper is outlined as follows.
 In
sect.\!\! 1, we present the hairy AdS black hole solution and study the VdW phase transition in the thermal entropy-temperature plane. In sect. 2, we employ the TCF  to probe the VdW phase transition. Especially, we  study the EAL  and critical exponent of the heat capacity in the framework of holography and find that the result  is similar as that obtained in the thermal entropy-temperature plane.
The conclusion and discussion is presented  in sect. 3.
Hereafter in this paper we use natural units ($G=c=\hbar=1$) for
simplicity.
\section{Thermodynamic phase transition of the hairy black hole}
The five-dimensional  hairy black hole solution can be written as[34]
\begin{eqnarray}\label{peturbation}
~ds^{2}=-f(r)dt^2+\frac{dr^2}{f(r)}+r^{2}[d\theta^2+\sin^2\theta (d\phi^2+\sin^2\theta d\psi^2)],
\end{eqnarray}
in which
\begin{eqnarray}
f(r)=\frac{e^2}{r^4}+\frac{r^2}{l^2}-\frac{m}{r^2}-\frac{q}{r^3}+1, 
\label{AADS}
\end{eqnarray}
where $e$ is the
electric charge, $m$ is the mass parameter, $l$ is the AdS radius that relates to the cosmology constant $\Lambda$,
and $q$ is related to the coupling constants of the
 conformal  field $b_0$, $b_1$, $b_2$. For the planar solution, $q=0$, while for the spherical symmetric black hole, $q$ can be expressed as
 \begin{eqnarray}
 q=\frac{64 \pi}{5} \varepsilon b_1 (\frac{-18 b_1}{5 b_0})^{3/2},
\end{eqnarray}
in which $ \varepsilon$  taking the values $-1,0,1$. In addition,  to satisfy the field equations,
the scalar coupling constants should obey the constraint
\begin{eqnarray}
100 b_0 b_2=9b_1^2.
\end{eqnarray}
As stressed in Ref.[34-35], the hair parameter $q$ is not a conserved charge corresponding to some symmetry, it  is a variable provided
the scalar field coupling constants are dynamic. In this paper, we will fix $q$ to investigate the phase structure of this black hole for $q$ has little effect on the  phase structure.

The black hole event horizon $r_h$ is the largest root of the equation  $f (r_h)=0$.~
At the event horizon, the Hawking temperature can be expressed as
\begin{equation}\label{tem}
T=\frac{-2 e^2 l^2+l^2 r_h \left(q+2 r_h^3\right)+4 r_h^6}{4 \pi  l^2 r_h^5},
\end{equation}
in which we have used the relation
\begin{equation}
m=\frac{e^2 l^2-l^2 q r_h+l^2 r_h^4+r_h^6}{l^2 r_h^2}.
 \end{equation}
In terms of the  AdS/CFT correspondence, the temperature in (\ref{tem}) can be treated as the temperature of the dual conformal field theory.
The Maxwell potential in this background is given by
\begin{eqnarray}
A_t=\frac{\sqrt{3 e}}{r_h^2}.
\end{eqnarray}
The entropy of  the black hole  is
\begin{equation}
S=\frac{\pi ^2 r_h^3}{2}-\frac{5 \pi ^2 q}{4}.\label{cp}
 \end{equation}
Substituting (\ref{cp}) into (\ref{tem}), we can get the following relation

\begin{equation}
\begin{split}
T=\frac{-2 \pi ^4 e^2 l^2+3\ 2^{2/3} \pi ^{10/3} l^2 q \sqrt[3]{5 \pi ^2 q+4 S}+2\ 2^{2/3} \pi ^{4/3} l^2 S \sqrt[3]{5 \pi ^2 q+4 S}+25 \pi ^4 q^2+40 \pi ^2 q S+16 S^2}{\sqrt[3]{2} l^2 \left(5 \pi ^3 q+4 \pi  S\right)^{5/3}}.\label{ts}
\end{split}
\end{equation}
Next, we will employ this equation to study the phase structure of the hairy black hole. The  Helmholtz free energy  of this  system is given by[34]
\begin{equation}
\begin{split}
F=M-TS=-\frac{5 \pi  e^2 q}{8 r_h^5}+\frac{5 \pi  e^2}{8 r_h^2}+\frac{5 \pi  q r_h}{4 l^2}-\frac{\pi  r_h^4}{8 l^2}+\frac{5 \pi  q^2}{16 r_h^4}+\frac{\pi  q}{8 r_h}+\frac{\pi  r_h^2}{8}, \label{tts}
\end{split}
\end{equation}
in which $M=\frac{3 \pi m}{8 G}$, where
 $G$ is the gravitational constant.

\begin{figure}[h!]%multicols环境下不能浮动，会导致图形或者表格丢失。
%只能当前位置（[]中参数必须是大写H），因此需要手动调位置
\centering
\includegraphics[scale=0.8]{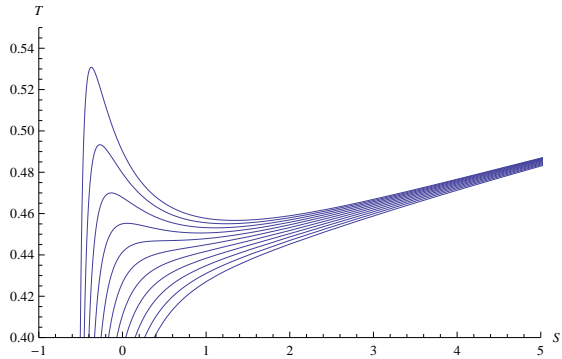}
\caption{\label{fig1} \small Relations between  temperature and thermal entropy for different $e$ with q = 0.05, curves from top to down correspond to  cases $e$ varies from 0.1 to 0.2 with step 0.01.
} %图题
\end{figure}

\begin{figure}[h!]%multicols环境下不能浮动，会导致图形或者表格丢失。
%只能当前位置（[]中参数必须是大写H），因此需要手动调位置
\centering
\includegraphics[scale=0.8]{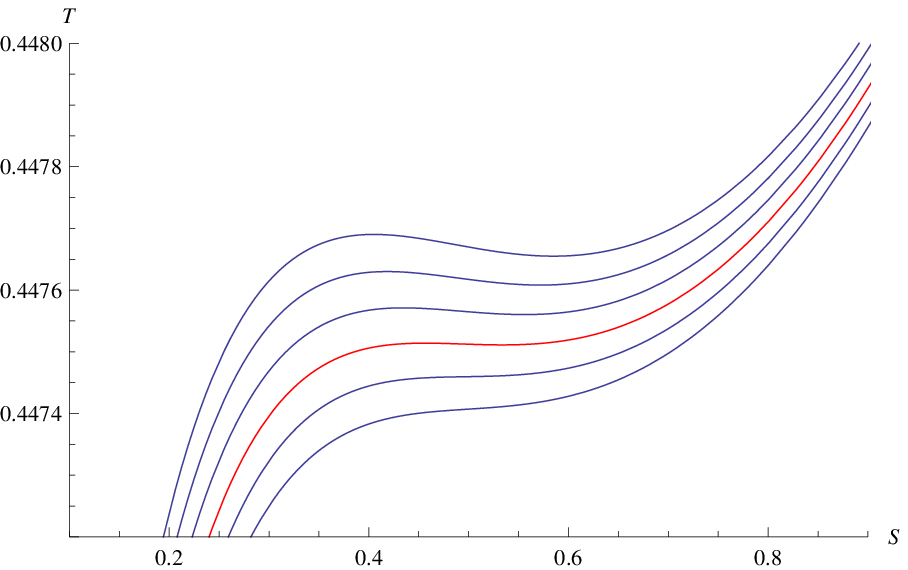}
\caption{\label{fig2} \small Relations between  temperature and thermal entropy  for different $e$ with q = 0.05, curves from top to down correspond to  cases $e$ varies from 0.1385 to 0.139 with step 0.0001.
} %图题
\end{figure}

 Now we concentrate  to study the phase structure of the hairy AdS black holes.  In fluid, the  VdW  phase transition is depicted in the  $P-V$ plane, where $P, V$ correspond to pressure and volume of fluid. In black holes,  there are two schemes to produce the VdW  phase transition. One is presented in Ref.[9-10] where the cosmology constant and curvature are treated  as pressure and volume. The other is presented in Ref.[8] in which one should adopt the following analogy
\begin{equation}\label{table1}
\begin{array}{|c|c|}
\hline
\multicolumn{2}{|c|}{\mbox{Analogy }} \\
\hline
\mbox{fluid} & \mbox{ AdS black hole} \\
\hline	
\mbox{temperature, T} &\mbox{charge, e} \\
\mbox{pressure, P} & \mbox{temperature, T( S,e )}\\
\mbox{volume, V} & \mbox{entropy, S}\\
\hline
\end{array}
\end{equation}
In this paper, we will follow the later scheme. In this case, the  phase structure depends on the charge of the black hole, and we know that there is a critical charge, for which the temperature  satisfies the following relation
 \begin{equation}
\left(\frac{\partial T}{\partial S}\right)_e=\left(\frac{\partial^2 T}{\partial S^2}\right)_e=0. \label{heat1}
 \end{equation}
In this paper, we will get the critical charge   numerically.  We will set $l=1$. We first plot a bunch of curves by taking different values of $e$ in the $T-S$ plane in Figure \ref{fig1}. From this figure, we can read off the rough critical value of the charge which satisfies the condition $(\frac{\partial T}{\partial S})_{e}=0$.  Having obtained this rough value, we
 plot several  curves in the $T-S$ plane further with smaller step so that we can get the probable critical value of $e$. From  Figure  \ref{fig2}, we find the probable critical value is $0.1388$, which is labeled by the red solid line. Finally, we adjust the value of $e$ by hand to find the exact value of $e$ that satisfies    $(\frac{\partial T}{\partial S})_{e}=0$ , which produces $e_c=0.13888$. With this critical value and  (\ref{heat1}), we can get the critical entropy $S_c=0.533926$.
Having obtained the critical charge and critical entropy, the critical temperature and critical free energy also can be produced by  the relations (\ref{ts}) and (\ref{tts}) directly.

\begin{figure}[h!]%multicols环境下不能浮动，会导致图形或者表格丢失。
%只能当前位置（[]中参数必须是大写H），因此需要手动调位置
\centering
\includegraphics[scale=0.8]{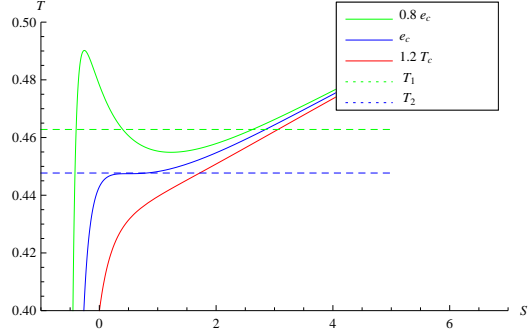}
\caption{\label{fig3} \small Relations between temperature and thermal entropy  for different $e$ with  $q=0.05$.
} %图题
\end{figure}

As the critical charge $e$ is given, we can set different charges to show the  VdW phase transition in the $T-S$ plane, which is shown in Figure \ref{fig3}.  From this figure, we know that as the value of the charge is smaller than the critical charge,   a small black hole, large black hole and an intermediate black hole coexist. There is a critical temperature, labeled as $T_1$. For the case $T<T_1$, the small black hole dominates while for the case $T>T_1$, the large black hole dominates.  The phase transition for the small black hole to the large black hole is first order. For the case $e=e_c$, we find the  unstable region vanishes and an inflection point emerges. The  heat capacity in this case is divergent, which implies that the phase transition is second order, the corresponding phase transition temperature is labeled as $T_2$.   For the case $e>e_c$,  the black hole is stable always.

\begin{figure}[h!]%multicols环境下不能浮动，会导致图形或者表格丢失。
%只能当前位置（[]中参数必须是大写H），因此需要手动调位置
\centering
\includegraphics[scale=0.8]{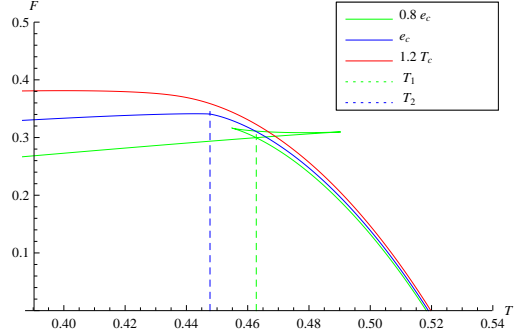}
\caption{\label{fig4} \small The $F-T$ relation for different $e$ with  $q=0.05$.
} %图题
\end{figure}

The phase structures can also be observed in the $F-T$ plane. From the green curve of Figure \ref{fig4}, we know the  swallowtail structure  corresponds to the unstable phase in the top curve of Figure \ref{fig3}. The
 value of the first order  phase transition temperature, $T_1=0.4628$, can be  read off from the horizontal coordinate of the junction.  From the blue curve of Figure \ref{fig4}, we also can read off the second order phase transition temperature $T_2=0.4477$.

The first order phase transition temperature $T_1$ also can be obtained from the EAL
\begin{equation}
\text{ } A_L\equiv\int_{S_1}^{S_3}T(S,e)dS\text{}=T_{1}(S_3-S_1)\text{}\equiv A_R,\text{  } \label{euqalarea}
 \end{equation}
in which $T(S,e)$ is the analytical function in (\ref{ts}), $S_1$ and  $S_3$ are the smallest and largest roots of the equation $T(S,e)=T_1$.
On the contrary, as $T_1$  is given, we can use it to check  the EAL  numerically. In fact, this relation holds always in thermodynamics. We give the numerical check here is to compare with the result which will be produced in the framework of holography in the next section.

For the case $q=0.05$, we know that $T_1=0.4628$, $S_1=0.40247$,  $S_3=2.62429$. Substituting these values into (\ref{euqalarea}), we find $A_L=1.01677$, $A_R=1.02826$. It obvious that $A_L$ and $A_R$ are equal nearly, which implies that the EAL   holds in the $T-S$ plane within our numerical accuracy.

For the second order phase transition, we know that near the critical temperature $T_2$, there is always a relation[11]
\begin{eqnarray}
\text{ }\log\mid T-T_2\mid \text{}\text{}=3 \log\mid S- S_c\mid +\text{constant}, \text{ }\text{ }   \label{cc2}
 \end{eqnarray}
 in which $S_c$ is the critical entropy corresponding the critical temperature $T_2$. With the definition of the heat capacity
 \begin{equation}
\text{ } C_{e}=T\frac{\partial S}{\partial T}\Big|_e \text{ }\label{capacity}.
 \end{equation}
 One can get further $C_e\sim(T-T_2)^{-2/3}$, namely the critical exponent is $-2/3$[11].
Next, we will check whether there is a similar linear relation  and critical exponent in the framework of holography.

\section{Probe the phase transition with two point correlation function}

\begin{figure}[h!]%multicols环境下不能浮动，会导致图形或者表格丢失。
%只能当前位置（[]中参数必须是大写H），因此需要手动调位置
\centering
\includegraphics[scale=0.9]{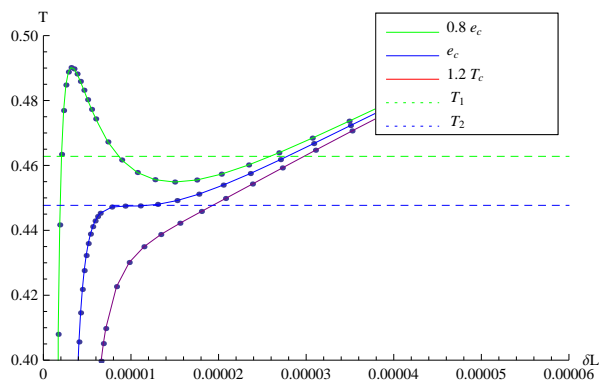}
\caption{\label{fig5} \small Relations between geodesic length and temperature  for different $e$ with  $q=0.05$, $\theta_0=0.1$.
} %图题
\end{figure}

\begin{figure}[h!]%multicols环境下不能浮动，会导致图形或者表格丢失。
%只能当前位置（[]中参数必须是大写H），因此需要手动调位置
\centering
\includegraphics[scale=0.9]{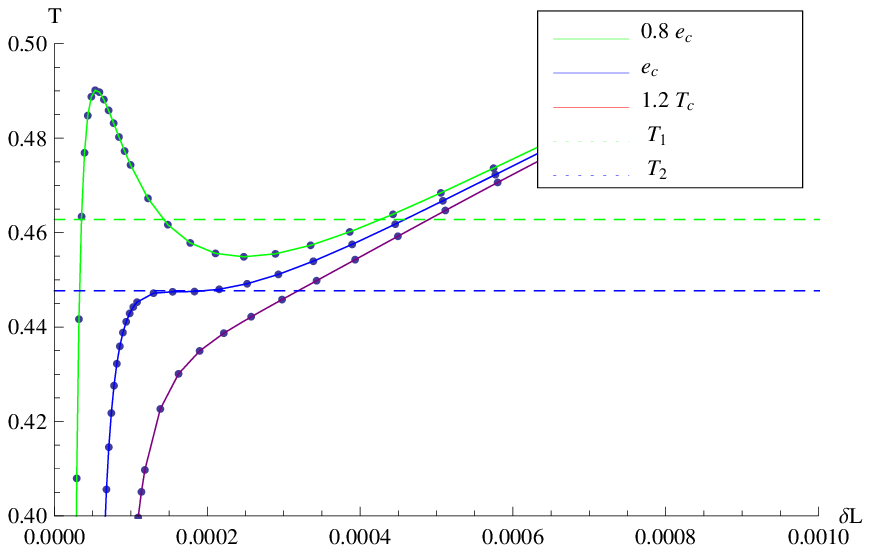}
\caption{\label{fig6} \small Relations between geodesic length and temperature  for different $e$ with  $q=0.05$, $\theta_0=0.2$.
} %图题
\end{figure}

Having obtained the thermodynamic phase structure of the hairy black hole, we will check whether this property can be probed by the TCF. In terms of the  AdS/CFT correspondence, we know that for the case that the  conformal
weight $\Delta$ is large,  the  equal time  two point function of operators  $\cal{O}$ can be computed by the length   of spacelike geodesics in the bulk geometry, that is  [36]
 \begin{equation}\label{llll4}
~\text{ }\big\langle {\cal{O}} (t,x_i) \text{}\text{}{\cal{O}}(t, x_j)\big\rangle  \approx
e^{-\Delta {\verb"L"}},~~ 
\end{equation}
in which $\verb"L"$ is the  length of the bulk geodesic connecting  two points $(t,
x_i)$ and $(t, x_j)$ on the AdS boundary.
For the spherically symmetric black hole,
  we will choose $(\phi=\frac{\pi}{2},\theta=0, \psi=0)$ and $(\phi=\frac{\pi}{2},\theta=\theta_0,\psi=\pi)$ as the two boundary points. Then with  $\theta$ to
    parameterize the trajectory, the proper length  can be written as
\begin{eqnarray}\label{rl}
~~~\verb"L"=\int_0 ^{\theta_0}\sqrt{\frac{{r}^{\prime 2}}{f(r)}+r^2} d\theta,~~
 \end{eqnarray}
 in which ${r}^{\prime}=dr/ d\theta$. Treating the integral as a lagrangian, we can get the motion equation of $r$, namely
 \begin{eqnarray}
~~~~\frac{1}{2} r(\theta ) r'(\theta )^2 g'(r(\theta ))-r(\theta ) g(r(\theta )) r''(\theta )+2 g(r(\theta )) r'(\theta )^2+r(\theta )^2 g(r(\theta ))^2=0.~~~
 \end{eqnarray}
With the following boundary conditions
\begin{eqnarray}
~~{r}^{\prime}(0)=0, r(0)= r_0,  \label{bon}
\end{eqnarray}
 we can
get the numeric result of $r(\theta)$ and further get $\verb"L"$ by substituting $r(\theta)$ into  (\ref{rl}).
As in [11], we  are interested in the regularized geodesic length as $\delta L\equiv \verb"L"-\verb"L"_0$, in which $\verb"L"_0$ is the geodesic length in pure AdS.
 As the value of $\delta L$ is given, we can get the  relation between  $\delta L$ and $T$  for different charge $e$. In this paper, we will also  explore whether this relation is  $\theta_0$  independent. Without loss of the generality, we choose $\theta_0=0.1, 0.2$ and set the corresponding  UV cutoff in the dual field
theory  to be $r(0.099)$, $r(0.199)$ respectively. The corresponding pictures are
  shown in Figure \ref{fig3}. It is obvious that both Figure \ref{fig5} and  Figure \ref{fig6} resemble as Figure \ref{fig3}  besides the scale of horizontal coordinate, which implies that the geodesic length owns the same phase structure as that of the thermal entropy.
To confirm this conclusion, we will   check the EAL  for the first order phase transition   and critical exponent for the second order phase transition in the $T-\delta L$ plane.

The EAL  in the $T-\delta L$ plane can be defined as
\begin{equation}
\text{}A_L\equiv\int_{\delta L_{min}}^{\delta L_{max}}T(\delta L,e)\text{}d\delta L=T_1\text{}(\delta L_{max}-\delta
L_{min})\equiv A_R \label{eeuqalarea},
 \end{equation}
in which  $T(\delta L,e)$  is an interpolating function obtained from the numeric data,  and $\delta L_{min}$,  $\delta L_{max}$ are the smallest and largest values of the  equation $T(\delta L,e)=T_{1}$.
For different $\theta_0$, the calculated results are listed in Table 1. From this table, we can see that for the unstable region of the first order phase transition in the $T-\delta L$ plane, the EAL  holds within our numeric accuracy. This conclusion is independent of the boundary separation $\theta_0$.

We also can investigate the critical exponent of the heat capacity for the second order phase transition in the $T-\delta L$  plane by   defining an analogous heat capacity
\begin{eqnarray}
C=T\frac{\partial \delta L}{\partial T}. \label{cheat2}
 \end{eqnarray}
Provided a similar relation as shown  in (\ref{cc2}) is satisfied, one can get the critical exponent of the analogous heat capacity immediately.

\begin{table}
\begin{center}\begin{tabular}{l|c|c|c|c|c|c|}
 %\MC{3}{c}{\text{caption}}\\[5pt]
 \hline
% & \multicolumn{3}{c||}{MGCDM}   & \multicolumn{3}{c}{$\Lambda$CDM}  \\ \hline
%                             &        MGCDM        &                  &             &      $\Lambda$CDM    &                   & \\ \hline
% \MC{3}{|c|c|}{\ZZ{-8pt}{15pt}\hfill\normalsize   \hfill  \hfill\normalsize MGCDM     \hfill\normalsize $\Lambda$CDM  }\\ \hline
% \ZZ{-6pt}{22pt}
    $\textrm{boundaery  size
}$                      &$T_{1}$ &     $\delta L_{min}$ &   $\delta L_{max}$     &$A_L$ &    $A_R$ &    $\textrm{relative ~error}$  \\  \hline
$\theta_0=0.1$        & 0.4628  &$ 0.0000014$   &$0.0000259$ & $ 0.0000111$ &   $0.0000113 $ &      1.8869\%\\ \hline
$\theta_0=0.2$        & 0.4628  &$0.0000354$   &$0.0004267$ & $ 0.0001812$ &   $0.0001811 $ &      0.4549\% \\ \hline
\end{tabular}
\end{center}
\caption{EAL  in the $T-\delta L$ plane, here relative ~error=${(A_L-A_R)/ A_R}$.}\label{tab2}
\end{table}

\begin{figure}[h!]
\centering
\includegraphics[scale=0.75]{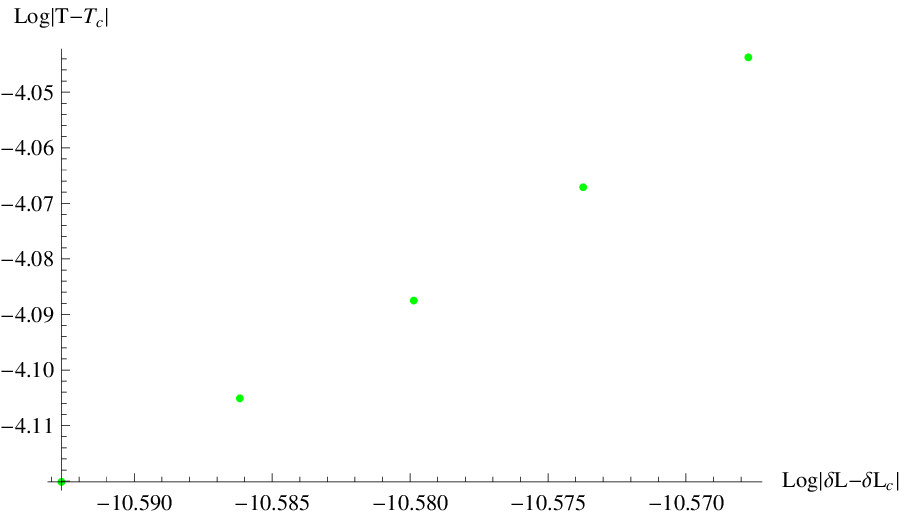}
\includegraphics[scale=0.75]{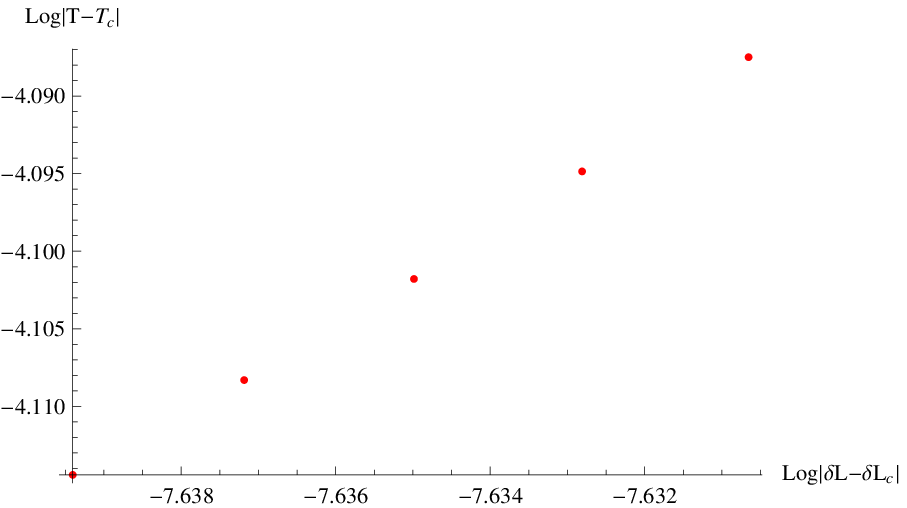}
\caption{\small \label{fig7}  Relations between $\log\mid T-T_2\mid$ and  $\log\mid\delta L-\delta L_c\mid $  for the case $q=0.05,e=e_c$. The left subgraph corresponds to the case $\theta_0=0.1$ while the right subgraph corresponds to $\theta_0=0.2$.  }
\end{figure}

So next, we are interested in the  relation between $ \log\mid T -T_2\mid$ and $\log\mid\delta L-\delta L_c\mid  $,
in which $\delta L_c$ is the solution that satisfies the relation  $T(\delta L,e)=T_{2}$.
For different $\theta_0$, the
numeric results are shown in Figure \ref{fig7}. By data fitting, the
straight lines  in  Figure \ref{fig7} can be fitted as

\begin{equation}
\log\mid T-T_2\mid=\begin{cases}
28.2901 + 3.05999  \log\mid\delta L-\delta L_c\mid,&$for$ ~q=0.05,e=e_c, \theta_0=0.1,\\
19.3767 + 3.07505 \log\mid\delta L-\delta L_c\mid, &$for$ ~ q=0.05,e=e_c, \theta_0=0.2.\\
\end{cases}
\end{equation}

 Obviously, for all the lines, the slope is about 3, which resembles as that in (\ref{cc2}). That is, the critical exponent of the analogous heat capacity  in $T-\delta L$ plane is the same as that in the $T-S$ plane, which once reinforce the conclusion that the TCF  can probe the phase structure of the hairy black hole.

\section{Conclusion and discussion}

As the usual charged black hole, we find the hairy black hole also exhibit the VdW phase transition. That is, the phase structure of the  hairy black hole depends on the charge of the spacetime. For the case that  the charge is smaller than the critical charge, the small black hole, immediate black hole, and large black hole coexist. The small black hole will transit to the large black hole as the temperature reaches to the critical temperature $T_1$. The order of this phase transition is first for the nonsmoothness of the free energy in the $F-T$ plane. As the charge equals to the critical charge, the immediate black hole vanishes and the order for the small black hole transit to the large black hole is second for the  heat capacity is divergent in this case. We also check the EAL  numerically for the first order phase transition and get the critical exponent of the heat capacity for the second order phase transition.

With the TCF , we also probe the phase structure of the black hole. For the TCF  in the field theory is dual to the geodesic length in the bulk, we thus employ the geodesic length to probe the phase structure of the hairy black hole. We find the phase structure in the $T-\delta L $ plane resembles as that in the $T-S $  plane, regardless of the size of the boundary separation. In addition, we find in the framework of holography, the EAL  still holds and the critical exponent of the analogous heat capacity is the same as that in the usual thermodynamic system.

In this paper, we fix the parameter $q=0.05$ to investigate the phase structure of the black hole. For other values of  $q$, we find the phase structure is similar as the case $q=0.05$. To avoid encumbrance, we will not list these values. In addition, we employ the analogy in  (\ref{table1}) to study the  VdW phase transition. In fact, as the cosmological constant is  treated as a thermodynamic pressure, $P$,  and its conjugate quantity as a thermodynamic volume, $V$, the VdW phase transition also can be constructed in the $P-V$ plane [13]. In [35], The holography entanglement entropy has been used to probe the phase structure in this case.  For the hairy black hole, the TCF  also can be used to probe its phase structure in  the $P-V$ plane.
\section*{Acknowledgements}

This work is supported  by the National Natural Science Foundation of China (Grant Nos. 11405016), China Postdoctoral Science Foundation (Grant No. 2016M590138), Natural Science
Foundation of  Education Committee of Chongqing (Grant No. KJ1500530), and Basic Research Project of Science and Technology Committee of Chongqing(Grant No. cstc2016jcyja0364).

\end{document}